\documentclass[aps,prl,amsmath,amssymb,reprint,superscriptaddress,twocolumn,showpacs]{revtex4-1}

 \usepackage{graphicx}
 \usepackage{amsmath}
 \usepackage{amsfonts}
 \usepackage{amssymb}
 \usepackage{pstricks}
 \usepackage{psfrag}
 \usepackage{epsfig}
 \usepackage{changes}
 \usepackage[ansinew]{inputenc}
\usepackage{mathptmx}
\usepackage{helvet}
\usepackage{textcomp}
\usepackage{ulem}

\begin{document}


\title{Predicting optimal hematocrit in silico}



\author{Alexander Farutin}
\affiliation{Univ. Grenoble Alpes, LIPHY, F-38000 Grenoble, France}
\affiliation{CNRS, LIPHY, F-38000 Grenoble, France}
\author{Zaiyi Shen}
\affiliation{Univ. Grenoble Alpes, LIPHY, F-38000 Grenoble, France}
\affiliation{CNRS, LIPHY, F-38000 Grenoble, France}
\author{Gael Prado}
\affiliation{Univ. Grenoble Alpes, LIPHY, F-38000 Grenoble, France}
\affiliation{CNRS, LIPHY, F-38000 Grenoble, France}
\author{Vassanti Audemar}
\affiliation{Univ. Grenoble Alpes, LIPHY, F-38000 Grenoble, France}
\affiliation{CNRS, LIPHY, F-38000 Grenoble, France}
\author{Hamid Ez-Zahraouy}
\affiliation{Laboratoire de Mati\`ere Condens\`ee et Sciences Interdisciplinaires, Faculty of Sciences, Mohammed V University of Rabat, Morocco}
\author{Abdelilah Benyoussef}
\affiliation{Laboratoire de Mati\`ere Condens\`ee et Sciences Interdisciplinaires, Faculty of Sciences, Mohammed V University of Rabat, Morocco}
\author{Benoit Polack}
\affiliation{Laboratoire d'H\'ematologie, CHU, Grenoble, France}
\affiliation{TIMC-IMAG/TheREx, CNRS UMR5525, Universit\'e Grenoble Alpes, Grenoble, France}
\author{Jens Harting}
\affiliation{Helmholtz Institute Erlangen-N\"urnberg for Renewable Energy (IEK-11), Forschungszentrum J\"ulich, F\"urther Strasse 248, 90429 N\"urnberg, Germany}
\affiliation{Department of Applied Physics, Eindhoven University of Technology, P.O. Box 513, 5600MB Eindhoven, The Netherlands}
\affiliation{Faculty of Science and Technology, Mesa+ Institute, University of Twente, 7500 AE Enschede, The Netherland}
\author{Petia M. Vlahovska}
\affiliation{Engineering Sciences and Applied Math, Northwestern University, Evanston 60208, USA}
\author{Thomas Podgorski}
\affiliation{Univ. Grenoble Alpes, LIPHY, F-38000 Grenoble, France}
\affiliation{CNRS, LIPHY, F-38000 Grenoble, France}
\author{Gwennou Coupier}
\affiliation{Univ. Grenoble Alpes, LIPHY, F-38000 Grenoble, France}
\affiliation{CNRS, LIPHY, F-38000 Grenoble, France}
\author{Chaouqi Misbah}
\email[]{chaouqi.misbah@univ-grenoble-alpes.fr}
\affiliation{Univ. Grenoble Alpes, LIPHY, F-38000 Grenoble, France}
\affiliation{CNRS, LIPHY, F-38000 Grenoble, France}


\date{\today}

\begin{abstract}
Optimal hematocrit $H_o$ maximizes oxygen transport.  In healthy humans,  the average hematocrit $H$ is in the range of 40-45$\%$, but it can significantly change in blood pathologies such as severe anemia  (low $H$) and polycythemia (high $H$). Whether the hematocrit level in humans corresponds to the optimal one is a long standing physiological question.
Here, using numerical simulations with the Lattice Boltzmann method and two mechanical models of the red blood cell (RBC) we predict the optimal hematocrit, and  explore  how altering the mechanical properties of RBCs affects $H_o$. We develop a simplified analytical theory that accounts for results obtained from numerical simulations and provides  insight into the physical mechanisms determining $H_o$. Our numerical and analytical models can easily be modified to incorporate a wide range of mechanical properties of RBCs as well as other  soft particles thereby providing means for the rational design of blood substitutes. Our work lays the foundations for systematic theoretical study of the  optimal hematocrit and its link with  pathological RBCs associated with various diseases (e.g. sickle cell anemia, diabetes mellitus, malaria, elliptocytosis).
\end{abstract}


\maketitle


\section{Introduction}
Oxygen delivery to tissues is provided by red blood cells (RBCs), which make up about $40-45\%$ of  blood volume (in the macrocirculation).
A long standing question is
the existence  of an optimal hematocrit (optimal volume percentage of RBCs), i.e.,  a hematocrit level above or below which
oxygen supply is impaired \cite{Hedrick1986,Linderkamp1992,Wells1991,Stark2012,Birchard1997,Barbee1971,Lipowsky1980a,shepherd82}.
A hematocrit maximizing oxygen transport  (i.e. a  maximum in the dependence of the RBC flow rate on  hematocrit) is  intuitively expected based on the following argument: while increasing the hematocrit increases the flux of oxygen-carrying RBCs,
it reduces blood flow rate because of increased flow resistance. Yet, the value of the optimal hematocrit  and the mechanisms controlling it are not known even though, given its medical importance, e.g., in blood transfusion, the topic has been extensively discussed in the physiology and hemorheology literature \cite{shepherd82,Stark2012}. A very basic question, ``why is the normal systemic hematocrit in humans about $40-45 \%$?'',  lacks definite answer. Moreover, since the hematocrit varies in the circulation (e.g., it decreases in the microcirculation  \cite{Popel2005}), another open question is ``where in the circulation is the hematocrit optimal?''.
 Solving these problems in silico is  challenging and highly non-trivial because of the complex nature of blood flow: it involves  many deformable RBCs moving and interacting  in  blood vessels with complicated  geometry \cite{Abkarian2008,Petia_review,Guido2009b,Abreu2014,Li2013}. Recent advances in computational power and novel simulation approaches, however, are beginning to make the task tractable and  the bottom-up  simulations of blood flow (by explicitly accounting for   blood elements) are becoming realistic \cite{Li2013}.
 Computational approaches based on continuum formulations \cite{Kraus1996,Cantat1999b,Pozrikidis2005,Bagchi2007,Dodson2008,Lac2007,Veerapaneni2009,Le2010,Zhao2010,Boedec2011,Salac2011,Lai2,Farutin2014b,rivera16} or particle-based models \cite{Noguchi2005a,Dupin2007,Clausen2010,Fedosov2011,Kaoui2011,Krueger2013, Pivkin2008} have  been   successfully applied to  simulations of the flow of single and many RBCs. These simulations have revealed that many  phenomena in blood flow such as platelet and leukocyte margination, plasma skimming, clustering \cite{Chien1987,Pries1992,Popel2005,Pries1996,McWhirter2009,Tomaiuolo2012,rivera16} have a purely mechanistic origin. For example, plasma skimming  and  hematocrit decrease in the microcirculation (F{\aa}hr{\ae}us-Lindquist  effect) \cite{Popel2005} arise from a cross-streamline migration of RBCs due to local flow perturbation by the deformable RBC  \cite{Olla1997,Cantat1999b,seifert1999}. Many experiments have pointed to and continue to reveal plethora of effects on the individual  and collective behaviors of RBCs under flow \cite{Fischer894,Dupire:2012,Guido2009b,Grandchamp2013,Lanotte:2016,SHEN201640}.

In this work, we study the RBC flow rate as a function of hematocrit, vessel size, and mechanical properties of RBCs.
We investigate this problem using both numerical simulations and analytical studies covering a wide hematocrit range. In the simulations we use two models for RBCs: (i) a 3D model accounting for both bending energy and in-plane shear elasticity, associated with the cytoskeleton (a network of proteins lying beneath the RBC membrane), and solved by the Lattice Boltzmann method (LBM), (ii) a 2D model based on the LBM   which accounts for membrane bending modes that shows the same trends, offering a faster tool to explore in the future wider ranges of parameters.
We find that optimality cannot be reached in all the vessels of the circulatory system at the same time and that for straight channels
the optimal hematocrit matches the actual hematocrit in the arterioles. For a vessel network (mimicking the real vasculature), optimality is reached between  small arteries, about 1 mm of diameter, down to large arterioles, in the range 100-200 $\mu$m (for humans the arteriole range  is $10-200 \mu m$\cite{Klabunde2012}).  Interestingly, for networks, the optimum occurs at the junction between macrocirculation (small arteries)  and microcirculations (arterioles). This is interesting as a major proportion of oxygen consumption  is known to take place in arterioles \cite{Gellis}, before reaching capillaries.

\section{Results}

 \begin{figure}[t]
\center{\includegraphics[width=0.45\textwidth]{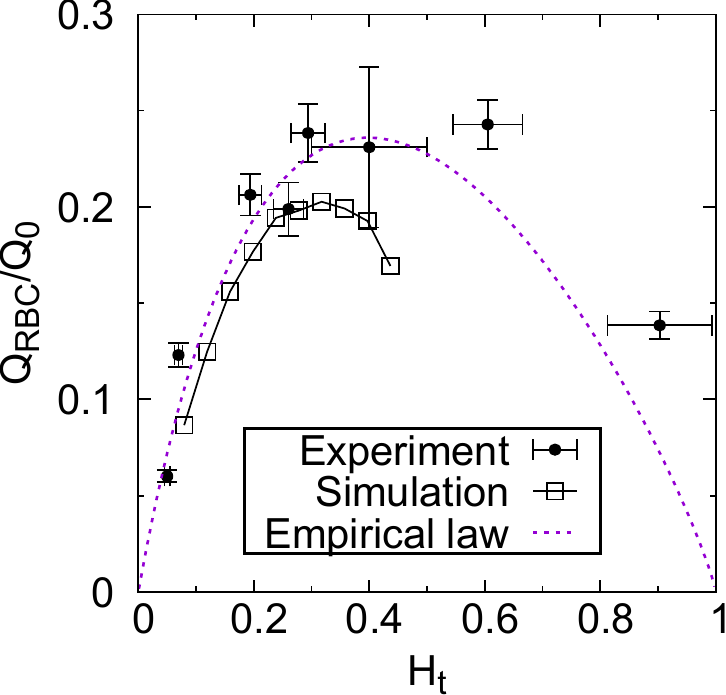}}
\caption{\label{fig1} RBC flow rate $Q_{RBC}/Q_0$ as a function of hematocrit $H_t$ for a channel width $W=40 \mu$m, according to experiment,  3D simulations, and empirical Pries model.  In the simulations, $C_s=0.18$, the channel width is $14.8 \;R $,  the channel length along the flow direction is about $11.7\; R$, and is equal to $6.2\; R $ in the orthogonal direction.}
\end{figure}

 \begin{figure}[t]
\includegraphics[width=0.45\textwidth]{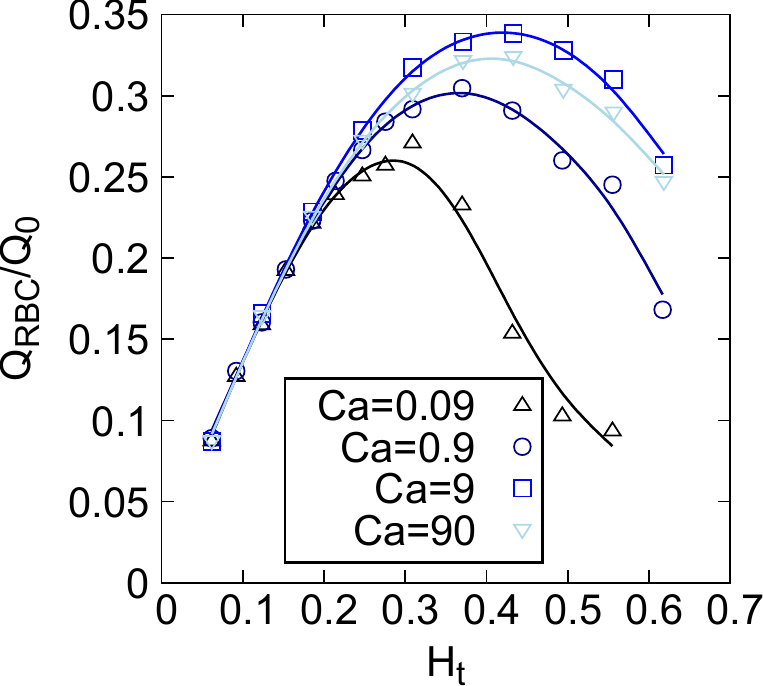}
\caption{\label{fig2}   Normalized RBC flow rate as a function of $C_a$ for a channel width $W=20 \mu$m. Reduced area is $\nu_{2D}=0.7.$}
\end{figure}

\begin{figure}[t]
\includegraphics[width=0.45\textwidth]{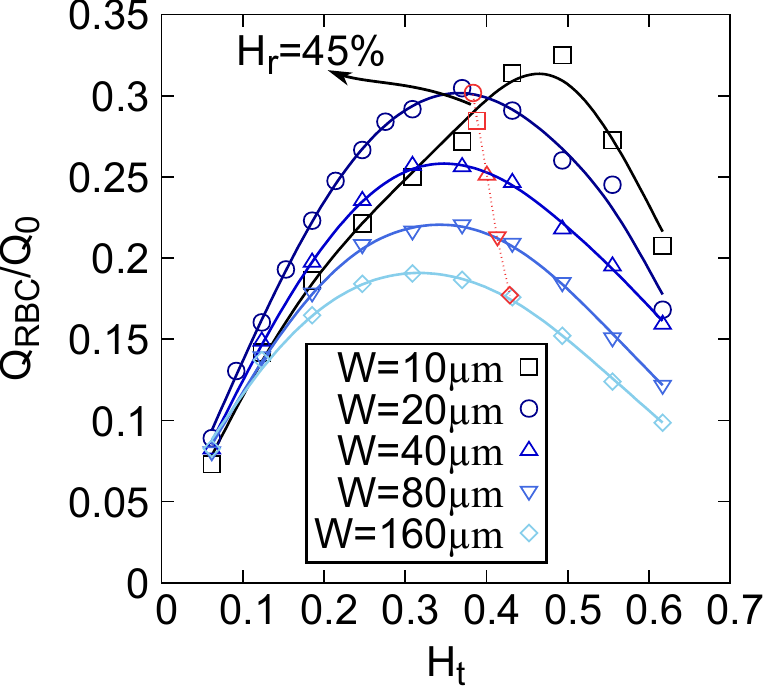}
\caption{\label{fig3} Normalized RBC flow rate as a function of hematocrit for channels with different widths. The red line indicates the $H_t$ value corresponding to a reservoir hematocrit of $45\%$.
$C_a=0.9$, and is  close to the values encountered in human arterioles. Reduced area is $\nu_{2D}=0.7.$}
\end{figure}

We investigate  first the existence of an optimal hematocrit  using numerical simulations with the LBM in 2D for computational efficiency, however, we also performed simulations in  3D to confirm that the behavior in 3D is qualitatively similar to that in 2D. We simulate the flow in a straight channel of length $L$ and width $W$ for several values of hematocrit $H_t$. The   channel widths range from $W=2R$ (capillary scale), where $R$ is the RBC effective radius,  up to $W=64 R$ (arteriole scale). The capillary number ranges from 0.09 to 90 in 2D and from 0.2 to 0.9 in 3D. The capillary number is defined in 2D and 3D as $C_{a}= \eta \langle \dot{\gamma}\rangle R^3 /\kappa$ and $C_s= \eta \langle\dot{\gamma}\rangle R /\mu _s$, respectively.  The average shear rate $\langle\dot{\gamma}\rangle$ is  defined by $2U_{max}/W$ ($U_{max}$ is the maximal velocity in the absence of red blood cells), $\kappa$ and $\mu_s$  are the bending and shear moduli of the RBC membrane, respectively. In 3D, a  capillary number based on shear elasticity can be estimated. Taking $R\sim 4\mu m$, $\mu_s= 4\mu$ N/m \cite{Suresh2006}, typical vessel diameters of human capillaries and arterioles (about 5-10 $\mu$m and 10-200 $\mu$m respectively \cite{Klabunde2012})   and typical maximal velocities in capillaries and arterioles ($1-10 $mm/s) \cite{KOUTSIARIS2010202} one finds an approximate range  $C_s=0.1-1$. The reduced volume $\nu$ in 3D and the reduced area $\nu_{2D}$ in 2D (see material and method for their definition) are varied from about 0.6 (healthy RBC value) up to 1 (sphere or circle) in order to explore their influence, as several diseases (spherocytosis, ellipsocytosis)  correspond to high value of $\nu$, close to one.

A validation experiment is performed with RBCs extracted from whole blood, and subject to a constant pressure difference. Unlike previous experimental studies \cite{Hedrick1986,Linderkamp1992} the hematocrit within the channel is extracted directly by light absorption techniques. This hematocrit is varied between 0.05 and 0.9 and the capillary number is estimated to be $C_s=0.01-0.19$. The microfluidic channel has rectangular cross section with a thickness of $40\; \mu$m much smaller than the other two dimensions.

Along this paper, we shall also refer  to the predictions made based on the empirical expression of healthy blood viscosity [denoted $\eta_{\mbox{p}}(H_t)$], obtained by  Pries et al. by data fitting \cite{Pries1992} (see SI).

Oxygen delivery is quantified  by the RBC flow rate, $Q_{RBC}$ normalized by the flow rate of the cell-free fluid $Q_0$ under same pressure gradient.  In the simulations, the  flow rate $Q_{RBC}$ is calculated by counting the number of RBCs that cross a given section in the channel, and after averaging over long runs. The determination of $Q_{RBC}$ in experiments is described in SI.
Figure \ref{fig1} shows that   $Q_{RBC}/Q_0$ exhibits a maximum as a function of hematocrit $H_t$. Similar results are found  in experiments and with the Pries et al. model \cite{Pries1992}, with an optimal  hematocrit about $0.4$.  In 3D simulations the optimum is around 0.3.

Next, we investigate this phenomenon in more detail, focusing on the effects of flow strength (measured by the capillary number) and channel width. The following simulations are performed in 2D.
Figure \ref{fig2}  shows that  the maximum of  $Q_{RBC}/Q_0$  is not very sensitive to the flow strength $C_a$: the optimal hematocrit  $H_o$ increases from 0.3 to about 0.4 over  3 orders of magnitude of increase in $C_a$. The slight increase in $H_o$ with $C_a$ arises from the increased RBC deformation (see below).   {{Interestingly, as $C_a$ approaches unity (values encountered in arterioles) $H_o$ quickly saturates to a value slightly larger than $0.4$. }}
 Figure \ref{fig3}  illustrates the effect of  channel width on the optimal hematocrit.   The results show that $H_o$ varies between 0.3 in large enough channels ($W=160\mu m$) to about 0.4 in smaller channels   ($W=10\mu m$).

\subsection{Optimality occurs in the arteriole range for straight channels}

In reality, the usual reference hematocrit of 0.4 and 0.45  for women and men respectively corresponds to values encountered in macrocirculation (usually measured in medical blood tests). It will be referred to as the reservoir hematocrit $H_r$. This value decreases in the microcirculation due (in particular) to the F\aa hr\ae us effect \cite{Fahraeus1929}.  This effect results from the tendency of RBCs to accumulate in the center of the vessel, acquiring thus a higher mean velocity than that of the plasma and leading to a diluted flow in small vessels: $H_t<H_r$.  In a given simulation the channel width and the number of cells are fixed, which corresponds to  prescribing the tube hematocrit $H_t$. If $\langle V_{RBC}\rangle$ denotes the mean velocity of RBC and $\langle V_T\rangle$ that of the whole suspension (RBCs plus the suspending fluid) in the tube, the total flow rate of RBCs across a section $A$ of the tube is given by $Q_{RBC}=\langle V_{RBC}\rangle A H_t$ and that of the whole  suspension is given by $ Q_T=\langle V_T\rangle A$. The ratio of the $Q_{RBC}$ to $Q_T$ provides the volume fraction of RBCs (or hematocrit) found in large vessels at the exit, $H_r$ (usually called reservoir hematocrit). We have thus the following relation
\begin{equation}
\label{reservoir}
H_r=\frac{Q_{RBC}}{Q_T}= H_t \frac{\langle V_{RBC}\rangle}{\langle V_T\rangle}
\end{equation}
Thus, measuring $\langle V_{RBC}\rangle/\langle V_T\rangle$ in the simulation yields directly the reservoir hematocrit. For each  channel width, performing simulation with different $H_t$, we can extract the relation between $H_r$ and $H_t$ (see Fig. S1 in SI). In large vessels (macrocirculation) the RBCs are homogeneously distribuged within the vessel section so that ${\langle V_{RBC}\rangle\simeq \langle V_T\rangle}$, implying $H_t\simeq H_r$. In contrast, in microcirculation the tendency of RBCs to migrate towards the vessel center means that  $\langle V_{RBC}\rangle > \langle V_T\rangle$, implying that $H_t < H_r$. 
 The red points in Figure \ref{fig3} show the tube hematocrit that would result from $H_r=0.45$ reservoir (systemic) hematocrit. The tube hematocrit $H_t$ is superoptimal in large enough channels (the red diamond in  Figure \ref{fig3} for $W=160 \; \mu $m and for larger channel widths -- not shown here) and approaches optimality for  $W=40\mu m$ and $W=20 \; \mu m$ (red upper triangle and red circle), which corresponds to the small-to-medium diameter range of human arterioles. This is interesting inasmuch as two thirds of the oxygen is  known to be delivered in the arteriolar   trees \cite{Gellis} before reaching capillaries. This picture is also confirmed (see below) by using a simplified model, and by using the Pries empirical model (see Fig. \ref{figp}), albeit with an optimum occurring for a slightly larger vessel diameter  (around $60 \mu m$).  This value, as well as the optimal one obtained in the simulations (Fig. \ref{fig3}), lies within  the arteriole range. On the contrary, in capillaries, the tube hematocrit is slightly suboptimal  for a reservoir hematocrit of 45\%.

  \subsection{Analytical model}
We develop an analytical model that captures the computational and experimental observations for the dependence of $Q_{RBC}$ on tube hematocrit $H_t$. This analytical model provides means for a quick estimate of the optimal hematocrit and insight into the mechanisms that control the phenomenon.
We adopt the two-fluid model for  blood flow: an inner core containing a homogeneous suspension of volume fraction $H$ surrounded by a cell-free layer of thickness $e$, flowing in a tube of radius  $R_0=W/2$. For given tube hematocrit $H_t$, $ H= H_t \times (\frac{R_0}{R_0-e})^2$.

The RBC flux is
\begin{equation}
Q_{RBC}(H_t)= H \int_0^{R_0-e} v (H,r) 2 \pi r  dr,\end{equation}
and the total flux is   given by
\begin{equation}
Q_{T}(H)= \int_0^{R_0}  v (H,r) 2 \pi r  dr
\end{equation}
 For a given pressure gradient (flow strength), the Stokes velocity profiles $v(H,r)$ in the outer cell-free annulus and in the core can be evaluated. 
  Assuming that the core fluid is a homogeneous, dense suspension of concentration $H$, its relative viscosity $\bar\eta\equiv \eta(H)/\eta_0$  can be estimated  from the  Krieger-Dougherty \cite{Krieger1959}  law  $\bar\eta=(1-H/H_m)^{-[\eta] H_m}$, where the maximum fraction $H_m=1$ (due to the fluid character of the the RBC membrane and its flexibility),  $[\eta]$ is the intrinsic viscosity, which is taken to be equal to 2.5, as for rigid spheres, in this simplistic model, and $\eta_0$ is the viscosity of the cell-free plasma.
 The thickness of the near-wall cell-free layer $e$ decreases linearly as hematocrit increases  \cite{fedosov10,shen16,rivera16}: $e=e_0(1-H_t/H_m)$ where $e_0$ is a constant characterizing the cell-free layer at low hematocrit (taken as   $e_0=2 \mu m$).
 Finally,
\begin{equation}
\begin{aligned}
\frac{Q_{RBC}}{Q_0} &= \frac{Q_{RBC}}{Q_T(H=0)} \\
&=\frac{H \big(R_0-e\big)^2\Big(R_0^2+\big(2 \bar\eta (H)-1\big)(2 eR_0-e^2)\Big)}{\bar\eta (H) R_0^4},
\end{aligned}
\end{equation}
where $e$ and $H$ are the functions of $H_t$ given above.

Figure  \ref{fig4}  illustrates the  prediction of this reduced model.
  As in simulations, it is seen that the optimal hematocrit increases with the confinement, and so does the corresponding RBC flux.  Had we assumed a fixed cell-free layer, independent of $H_t$, we would then have missed the fact (data not shown, see also \cite{Stark2012}) that optimal hematocrit increases with confinement.  The red line corresponds to the location of the tube hematocrit if the reservoir hematocrit (i.e. the hematocrit in macrocirculation) is equal to 0.45. Here again we find, as in direct simulation (Figure \ref{fig3}), that the optimal hematocrit corresponds to a channel diameter of about $30\; \mu m$ (red upper triangle in Fig. \ref{fig4}) meaning that optimality is reached in the arterial tree, and not in capillaries.

\subsection{Optimal hematocrit in vascular network}

\begin{figure}
\centering
\includegraphics[width=0.45\textwidth]{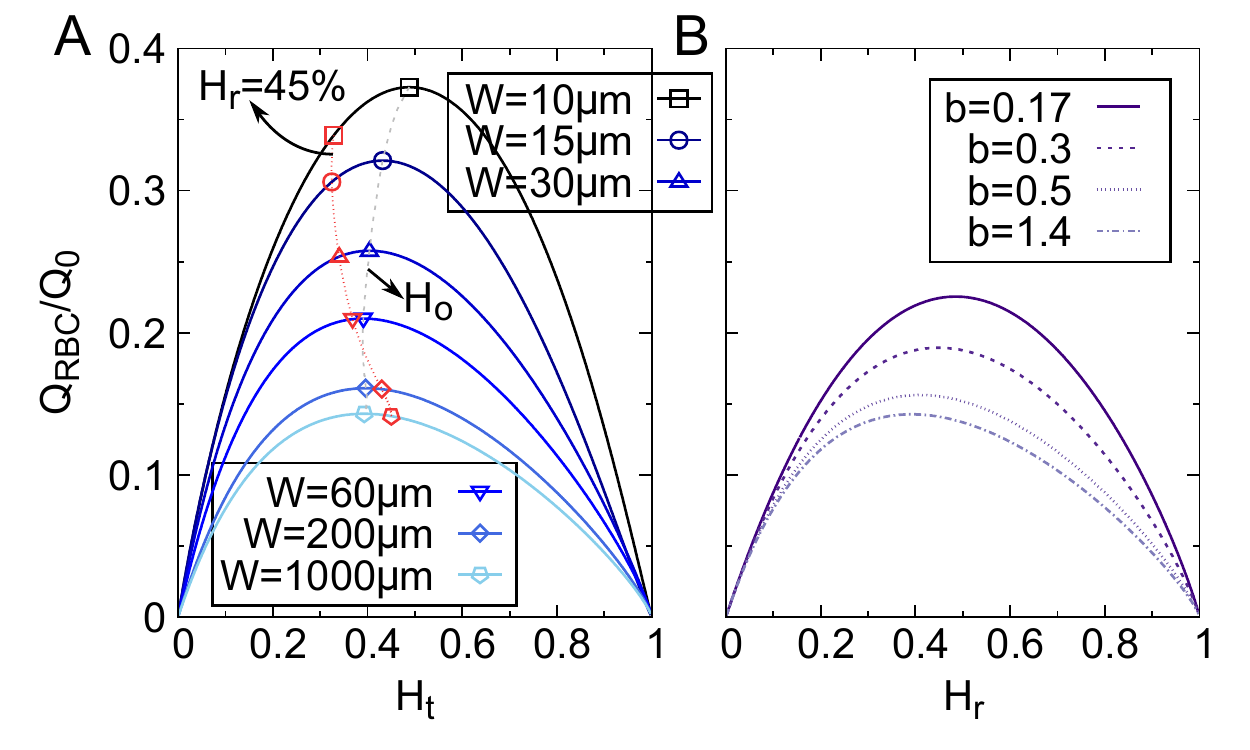}
\caption{\label{figp}A:  The RBC flow rate as a function of tube hematocrit for  channels of different diameters calculated
from Pries empirical model;
The red line indicates the $H_t$ value corresponding to a reservoir hematocrit of 45\%. B: The RBC flow rate for a model network for different values of $b$ extracted from literature. The curves are similar to those for large channels that would be obtained from panel A by considering $H_r$ instead of $H_t$. Small values of $b$ give more weight to the smallest vessels, hence an increase of the optimal hematocrit. Still, the contribution of large vessels is predominant.}
\end{figure}

According to the simulations (Fig.  \ref{fig3}) and the analytical results (Fig. \ref{fig4} and Fig. \ref{figp}A for a more quantitative analysis), when the reservoir hematocrit is equal to 0.45, the tube hamatocrit in microcirculation goes down to about 0.3 (see red lines in Fig. \ref{fig3} and Fig. \ref{figp}A). Experimental studies of tube flow \cite{Albrecht1979,Fahraeus1929,Barbee1971}
yielded minimal values for the ratio $H_t/H_r$ of about 0.7 for tubes of 20 $\mu m$ and 10 $\mu m$ diameters, which is consistent with this study which provides $0.3/0.45 \sim 0.66$. However, in vivo studies \cite{Sarelius1982,Lipowsky1980a,Klitzman1979} have shown that this ratio can fall in the range 0.3-0.5  in the microcirculation.  In vivo measurements have suggested that this is dependent on network topography and local flow rates. While, for that reason, our study can not be directly applied to a vascular network, we have attempted to analyze this issue on the basis of Pries empirical model (see Fig. \ref{figp}A) with the hope to get a more quantitative result.  We have considered the cases where the network consists of $N$ levels, where each tube of generation $i$ with radius $R(i)$ and length $L(i)$ divides symmetrically into two channels of level $i+1$  with radius $R(i+1)$ and length and $L(i+1)$ (see SI).
$R(i)^{1/a}=2 R(i+1)^{1/a}$ and $L(i)^{1/b}=2 L(i+1)^{1/b}$, where $a$ and $b$ are exponents that can be obtained by fitting in-vivo data \cite{newberry15}. This provides\cite{newberry15} the ranges $1/3-1/2$ and $0.17-1.4$ for $a$ and $b$, respectively.  We considered a network starting from a channel of about $1$ mm with successive bifurcations down to 5 $\mu m$. Taking $a=1/3$ (consistent with Murray's law \cite{Murray1926}) this analysis reveals  (Fig. \ref{figp}B) that the dominant contribution arises from largest vessels of the network. In other words, the overall efficiency of flow rate of RBCs could be  dictated by large vessels, despite the fact that oxygen delivery takes place in microcirculation. This would point to the fact that optimization of oxygen transport capacity may be regulated  upstream of the microcirculation, close to the transition zone between small arteries and arterioles. Caution is  however necessary since the situation is less clear for $a=1/2$, for instance. In addition, a continuum model may fail to describe the dynamics in small  vessels of the networks \cite{SHEN201640}.
 A systematic simulation taking into account the corpuscular nature of blood and the heterogeneities of networks are necessary before making general conclusive answers.

\subsection{Blood disorder}
\begin{figure}
\centering
\includegraphics[width=0.45\textwidth]{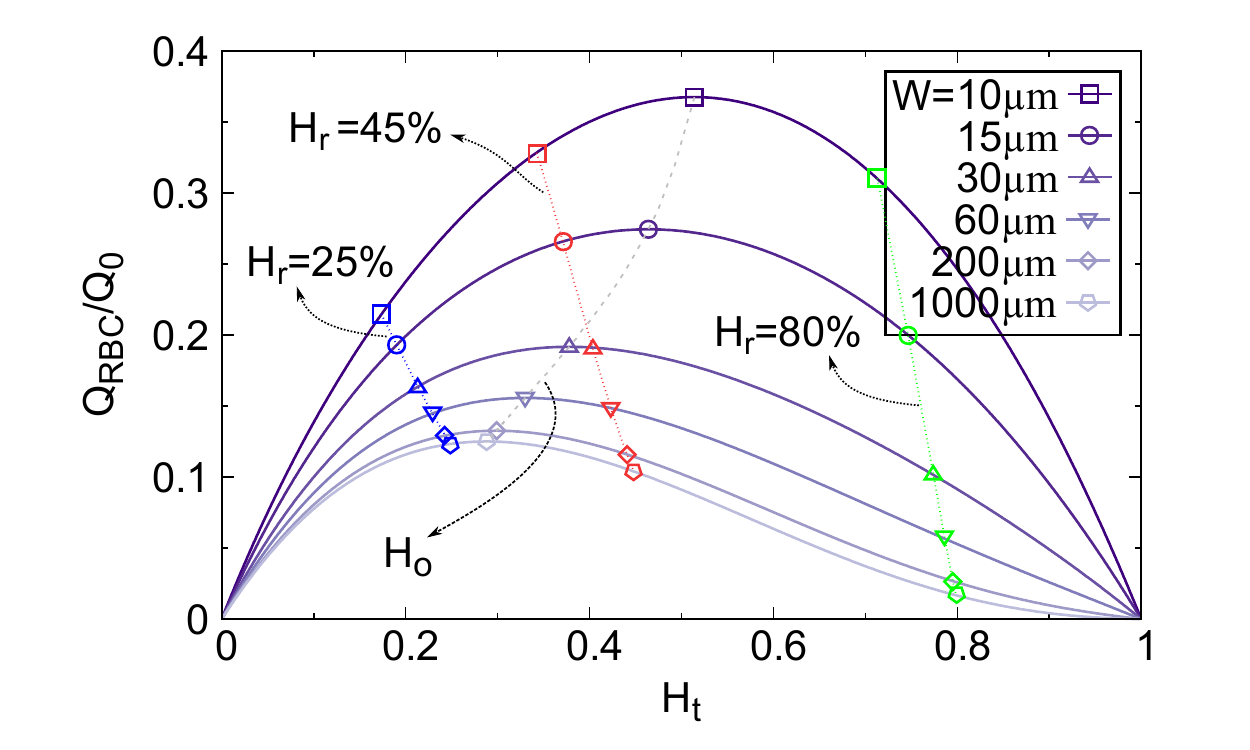}
\caption{\label{fig4} The RBC flow rate as a function of tube hematocrit for  channels with different diameter  calculated from our minimal model.
The red line indicates the $H_t$ value corresponding to a reservoir hematocrit $H_r$ of 45\%  for  the corresponding model. Blue and green lines correspond to high $H_r$ (80\% , such as in polycytemia), and low $H_r$ (25 \% for anemia).
}
\end{figure}

Blood disorders  are known to affect blood flow and rheology. Thus we have investigated the implication of some disorders for the oxygen transport capacity. The first example
corresponds to patients suffering from low or high level of hematocrit (anemia and polycythemia, respectively).
An outstanding feature emerges. Indeed,  if the reservoir hematocrit is small enough ($H_r=0.25$), as happens in severe anemia disease -- blue line in Fig. \ref{fig4} --, or large enough ($H_r=0.8$) as is the case in polycythemia vera -- green line in Fig. \ref{fig4} --, then the tube hematocrit does not cross any optimal value. In the anemia disease case the hematocrit remains suboptimal, while it remains superoptimal in the polycythemia disease. In the first case the attained flow rate in arterioles is  about 25\% lower than in healthy subjects, while the situation is more severe in the second case where the reduction can attain about 60\%.

\begin{figure*}
\centering
\includegraphics[width=0.9\textwidth]{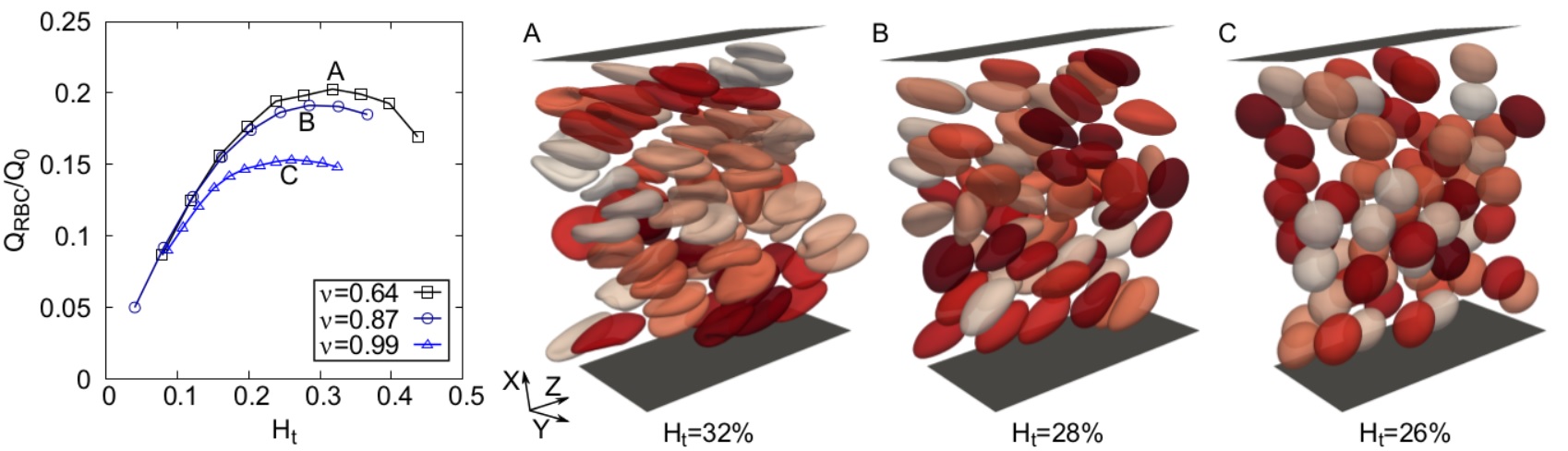}
\caption{\label{zaiyi3D} Left: The normalized RBC flow rate as a function of tube hematocrit for different reduced volumes.
Viscosity contrast is 1 and the channel width $W$ is $40 \mu m$ (other dimensions as in Fig. \ref{fig1}) and  $C_s=0.18$. A, B and C are snapshots of suspension configurations for the three  different reduced volumes shown for the corresponding optimal hematocrit.}
\end{figure*}

Other blood disorders are related to RBC shapes.
  For example, in  spherocytosis (a hereditary disease  caused by molecular defects in several proteins of the cytoskeleton that can lead to asthenia, polypnea, and an increased heart rate among other symptoms) RBCs shape is close to a sphere, which quantitatively corresponds to reduced volume close to 1. Here the simulation is performed both in 2D (data not shown, same conclusions) and in  3D  using the classical Skalak et al. model for the cytoskeleton \cite{Skalak_elas}. Figure \ref{zaiyi3D} shows that a spherical shape of the RBCs lowers both the optimal $H_o$ and the maximum RBC flow rate. For example, when the reduced volume $\nu$ passes from 0.64 (healthy RBC) to 0.94 (close enough to a spherical shape) the maximal carrying capacity drops by about 25 $\%$. The origin of this collapse is attributed to the fact that
RBCs are less deformable because of nearly spherical shape reducing their ability
to squeeze and accommodate high $H$. Moreover, cross-stream migration is suppressed and the cell-free layer diminishes leading to increased flow resistance. Both  effects (packing and cell-free layer thinning) contribute to a collapse of the oxygen carrying capacity. Those effects can be qualitatively included in our minimal model by modifying accordingly the rheology law for the core or the cell-free layer thickness. This leads to the same conclusion (see SI).

Finally, many pathologies are accompanied by an alteration of the mechanical properties (e.g. shear elasticity) of the RBC membrane. For example, in sickle cell and malaria diseases, the RBC elastic modulus can be significantly higher (up to about three times higher) than within healthy subjects. An increase of the elastic modulus is equivalent to a decrease of the capillary number. Our data in Fig. \ref{fig2} show that a reduction of the capillary number leads to a significant collapse of the RBC flow rate.

 \subsection{Stretchable capsules boost the oxygen carrying capacity}

\begin{figure}
\centering
\includegraphics[width=0.5\textwidth]{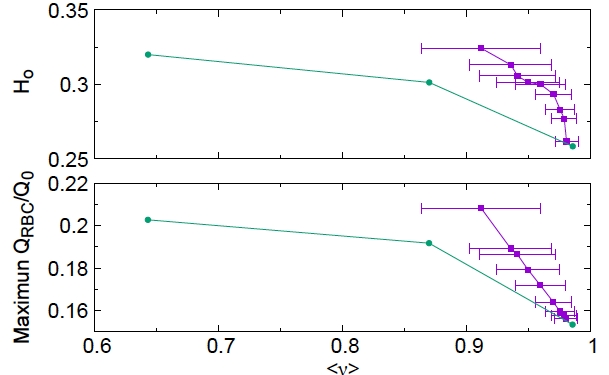}
\caption{\label{fig6}  Green line: Optimal flux and optimal hematocrit as a function of the actual reduced volume (3D simulations) for incompressible membranes. violet lines: the same simulation with stretchable  capsules. For each simulation the compressional elasticity has been varied in order to allow for different average reduced volume $\langle  \nu \rangle$. Each capsule experiences different shear stresses within the channel, and has thus different  surface area (and thus reduced volume); the horizontal bars provide the distribution width of reduced volumes within the same suspension. $C_s=0.18$, as in Fig. \ref{fig1} }\end{figure}

A major branch of research in the context of blood substitutes is directed towards
  hemoglobin encapsulation strategies.  It is put forward here that a slightly stretchable encapsulating membrane (like polymer-based capsules)
would lead
 to a significant  enhancement of  oxygen transport capacity. This idea emerged from
  a numerical analysis of  the role of membrane compressional elasticity on the maximum hemoglobin flow rate. Figure \ref{fig6} (violet data) shows the results; the green data correspond to incompressible membranes. In  each simulation we have selected  an initial reduced volume { $\nu=1$} (the same for all cells) and performed several simulations, each time with different elastic properties. Each cell within the same uniform suspension  (i.e.  all cells have the same properties)   experiences different shear stress and thus will be more or less stretched. The average actual reduced volume within the suspension is shown as a  filled square, whereas the horizontal bars show the distribution of reduced volume for each simulation (i.e.  for a given stretching elastic coefficient).
  One sees that a moderate extensibility, allowing an average decrease of the reduced volume of the capsules by as little as 10$\%$, leads to a maximum flow rate equivalent to that of  healthy conditions, even though the capsules are quasi-spherical. This result also implies that if the blood substitute elements were biconcave, a significantly lower hematocrit than $40-45 \%$ can provide  the same oxygen carriage capacity.   Membrane elasticity, even when moderate, emerges   thus as an important element for future  design of oxygen carrier materials.

\section{Conclusion}
Through intensive numerical 2D and 3D simulations we  provided information on the behavior of the RBC flow rate as a function of hematocrit, which is directly linked to the oxygen carriage capacity. Interestingly, the values obtained for optimal hematocrit in our  simulations for vessel sizes corresponding to macrocirculation and intermediate microcirculation (arterioles) are close enough to the corresponding physiologically admitted values.
Strong alterations are reported  if the reduced volume of RBCs is increased sufficiently, as is known for elliptocytosis and spherocytosis diseases. Not only is the flow rate of RBCs reduced in this case compared to the flow of healthy RBCs at the same hematocrit but also the optimal hematocrit is observed to be significantly lower. The lower value of RBC flow rate within  patients suffering  these diseases implies a severe collapse of oxygen delivery, which could lead to an increased heart load in order to maintain appropriate perfusion levels. It is known that elliptocytosis and spherocytosis diseases are accompanied by a reduction of the RBC count. This might represent a natural adaptation that actually improves oxygen delivery.
Stretchable membranes are found  to be a key element in boosting oxygen transport. This study provides a more precise framework for predicting the impact of several blood diseases on oxygen transport, and for {suggesting}  new paradigms in blood flow research. The information generated by this study may guide the development of new soft materials, such as blood substitutes, and advance the {tuning process} and optimization of oxygen carriers.

\section{Materials and Methods}


We represent RBCs as either a contour endowed with bending energy immersed in 2D fluid, or as a surface having both
bending energy and shear elasticity in a 3D fluid. The simulation uses a 2D and 3D Lattice Boltzmann method. The bending energy is given by $E=(\kappa/2) \int H^2 dA$ with $H$ the mean curvature, $dA$ the arclength in 2D or area in 3D, and $\kappa$ the bending rigidity modulus. In 3D, the membrane has  additionally a  shear elastic energy written as
$\mu_s(I_1^2 + 2I_1-I_2)/12+  \kappa_\alpha I_2^2/12$,
where $\mu_s$ is the shear elastic modulus and $\kappa_\alpha$ is the area dilation modulus.
$I_1$ and $I_2$ are the in-plane strain invariants (see \cite{Krueger2011}).
$\kappa_\alpha/\mu_s=200$ is chosen large enough to preserve membrane area locally.

We impose a pressure difference $\Delta p$ between the inlet and outlet of the channel. In the absence of RBCs the flow is of a Poiseuille type. In 2D {(3D)} the hematocrit is defined as $H_t= N A /{\cal A}$ ($N V/{\cal V})$  where $N$ is the number of cells in the channel, $A$  the area of cell and ${ V}$ the volume, whereas ${\cal A}$ and ${\cal V}$ designate the total area and volume of the channel. $\eta$ is taken as $1.2$ m.Pa.s (plasma viscosity), and the enclosed fluid within RBCs is taken to have the same viscosity. The precise values of the viscosity contrast have little influence on the results. We define the capillary number (which is a measure of the flow strength over the RBC's mechanical resistance) associated with bending and shear elastic modes as
$C_b= \eta \langle\dot{\gamma}\rangle R^3/\kappa$ and $C_{s}= \eta \langle\dot{\gamma}\rangle R/\mu_s$, with $\langle\dot{\gamma}\rangle=2U_{max}/W$,  where $U_{max}$ is the maximum velocity in the channel in the absence of cells. We have taken   $\kappa \simeq 3 \; 10^{-19}\;$ J and $\mu_s\simeq 4 $ \textmu N/m. We define (in 2D) the reduced area $\nu_{2D}\equiv (A/\pi)/(p/2\pi)^2$ (with $p$ the perimeter and  $A$ the enclosed area)
and the reduced volume (in 3D) $\nu\equiv [V/(4\pi/3)]/[A/4\pi]^{3/2}$.  For a healthy RBC $\nu \simeq 0.64$. We define in 2D $R =\sqrt{A/\pi}$, and in 3D $R=[3V/(4\pi)]^{1/3}$ as the typical RBC radius. For a healthy RBC we have $R\simeq 2.7 \mu m$.

For experiments RBCs were extracted from whole blood by successive washes
in PBS solution and centrifugation. After each centrifugation, the supernatant
was pipetted out and PBS was added to refill the tube. The washing/centrifugation cycle was repeated three times. After the RBCs were isolated, an adequate quantity was pipetted and suspended in a  buffer solution having the same density as RBCs  (137mL ultrapure water, 63mL Optiprep --- a iodixanol solution from Axis Shield ---, 1 tablet of PBS, 198mg glucose, 200mg BSA, density $\rho \approx 1.1\ $g$ \cdot $mL$^{-1}$) to produce samples of hematocrits ranging from $H_{r} \approx 0.05\ \text{to}\ 0.96$ with no sedimentation. Those reservoir hematocrits were measured using glass hematocrit tubes and centrifugation (of samples diluted in buffer solutions with no Optiprep). The tube hematocrit was determined by light absorption techniques (see SI). A given pressure difference was applied corresponding to  shear elasticity capillary numbers (the most relevant one for RBC) in the range  $C_s=0.01-0.19$. The microfluidic channel has a  rectangular cross section, with thickness $W=40\;\mu m$, whereas the two other dimensions are comparably large (see SI). The RBCs flow rate measurement is described in SI.


\begin{acknowledgments}
This work was partially supported by
CNES (Centre National d'Etudes Spatiales) and by the French-German university programme "Living Fluids" (grant CFDA-Q1-14). C.M. thanks CNRST (project FINCOME).
\end{acknowledgments}

\bibliography{varsaw_biblio}

\clearpage
\onecolumngrid
\newpage

\makeatletter 
\def\tagform@#1{\maketag@@@{(S\ignorespaces#1\unskip\@@italiccorr)}}
\makeatother

\makeatletter \renewcommand{\fnum@figure}
{\figurename~S\thefigure}
\makeatother

\setcounter{equation}{0}
\setcounter{figure}{0}

\begin{center}
  {\Large \bf Supplemental material for: Predicting optimal hematocrit in silico}
\end{center}

\medskip

\section{Experimental confirmation of the optimum}

We describe a straightforward determination of the blood flux at constant pressure drop in a confined geometry, which confirms the general trend that is observed in simulations.

\subsubsection*{Preparation of blood samples}

Blood samples from healthy donors were obtained through the \'Etablissement
Fran\c{c}ais du Sang (Grenoble, France) and stored at $4^{\circ}$C until use.
Red blood cells (RBCs) were extracted from whole blood by successive washes
in PBS solution and centrifugation. After each centrifugation, the supernatant
was pipetted out and PBS was added to refill the tube. The washing/centrifugation cycle was repeated three times. After the RBCs were isolated, an adequate quantity was pipetted and suspended in a  buffer solution having the same density as RBCs  (137mL ultrapure water, 63mL Optiprep --- a iodixanol solution from Axis Shield ---, 1 tablet of PBS, 198mg glucose, 200mg BSA, density $\rho \approx 1.1\ $g$ \cdot $mL$^{-1}$) to produce samples of hematocrits ranging from $H_{r} \approx 0.05\ \text{to}\ 0.96$ with no sedimentation. Those reservoir hematocrits were measured using glass hematocrit tubes and centrifugation (of samples diluted in buffer solutions with no Optiprep).

\subsubsection*{Determination of tube hematocrit}

Due to scattering, the absorption in a RBC suspension depends non-linearly on the hematocrit \cite{Lipowsky1980b}. A calibration procedure is required to measure the absorption coefficient at known hematocrits. For each sample of measured hematocrit $H_r$, as well as for the suspending iso-dense solution, a small drop is deposited at the center of a PDMS pool of same height as the channel, a glass coverslip is placed on top of the pool and its contour pressed to adhere to the glass. Images are then acquired at different positions to account for possible inhomogeneities. This provides a calibrated relationship between the absorption coefficient $-\log\left(\frac{I(H_r)}{I(H_r=0)}\right)$  and $H$, where $I(H_r)$ is the average light intensity in the image field for a suspension of hematocrit $H_r$ and $I(H_r=0)$ corresponds to the iso-dense solution.

In order to use the calibration curve, the optical parameters are kept strictly identical during the flow experiments (diaphragm aperture, lightning conditions, light condenser position, camera parameters). Additionally, we checked that the behaviour of the absorption coefficient was similar when using different blood samples, although due to the intrinsic variability in biological samples, it was necessary to repeat the calibration experiment for each blood sample.

\begin{figure}[h]
\centering
\includegraphics[width=9cm]{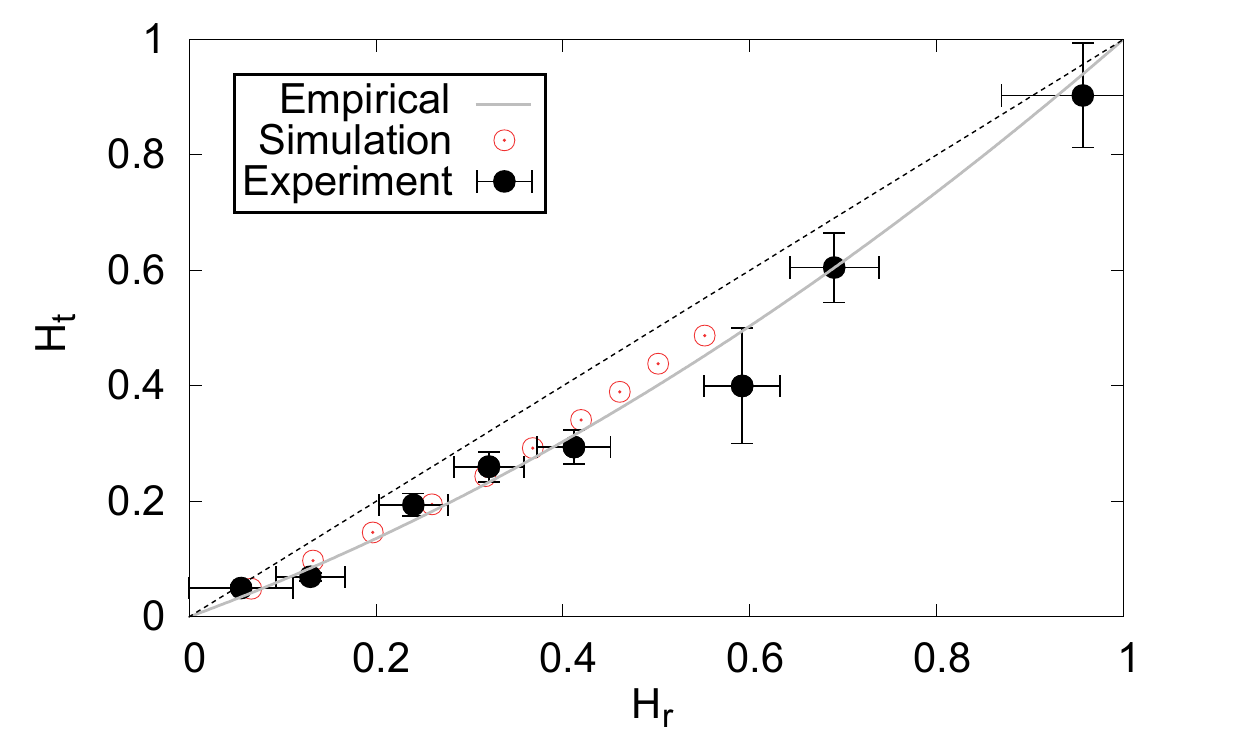}
\caption{\label{figHtvsH0}Black dots: Tube hematocrit $H_t$ determined by light absorption versus reservoir hematocrit $H_r$ in the flat channel of thickness 40 \textmu m. The open circles show the 3D numerical results. The dashed dark grey line is a guide to the eye of slope unity, the solid light grey line is the empirical relation from Ref. \cite{Pries1992} in a tube of diameter  $D=40$ \textmu m.}
\end{figure}

Interestingly, the relationship between $H_t$ and $H_r$  that we determined in the flat channel of thickness 40 \textmu m that we use in the following shows a good agreement with the empirical model of Pries \textit{et al.} \cite{pries92} that was proposed for a cylindrical tube of diamter 40 \textmu m (see Fig. S\ref{figHtvsH0}).
%

%

\subsubsection*{RBC flux}

\begin{figure}[h]
\centering
\includegraphics[width=8cm]{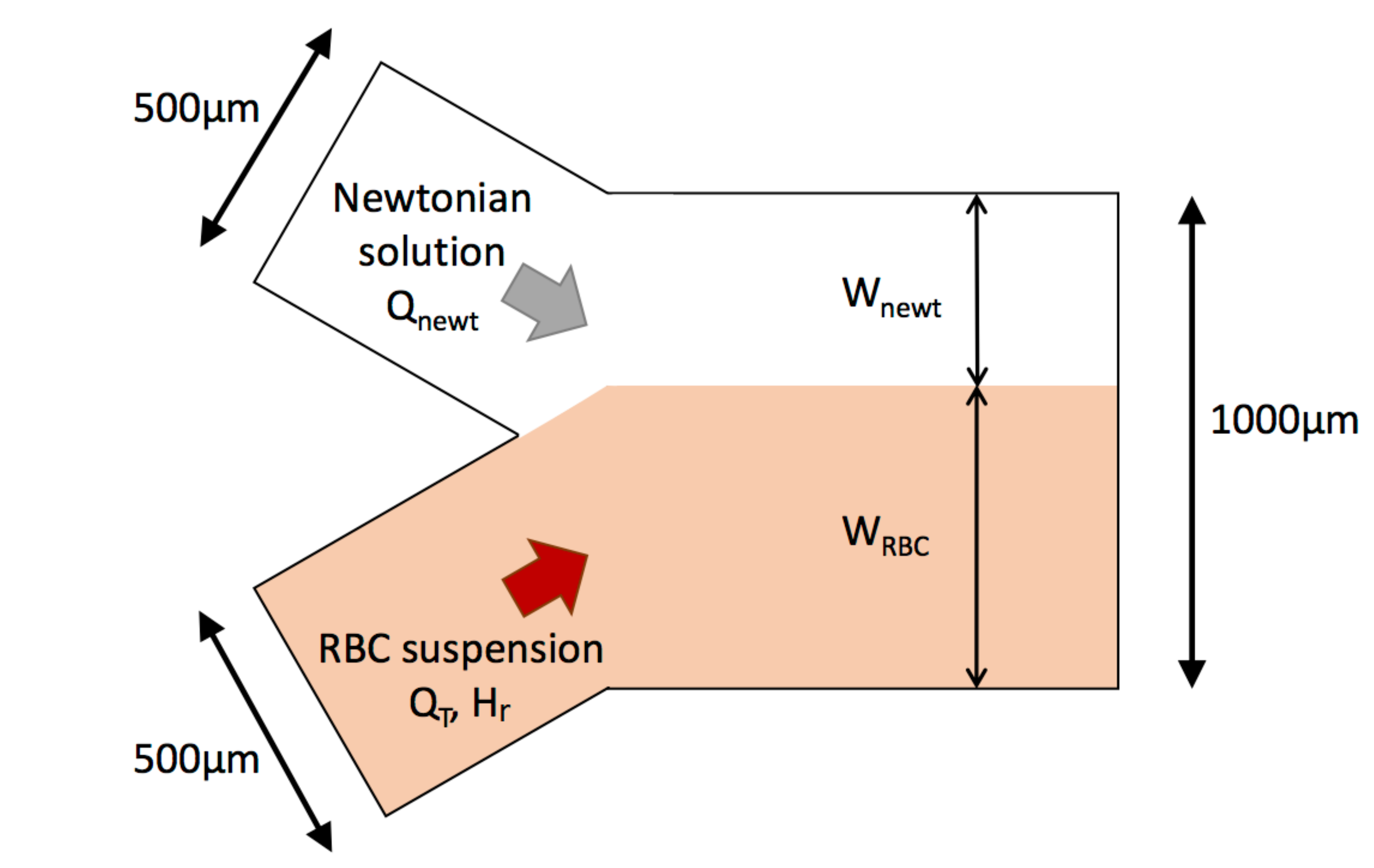}
\caption{\label{figFlowGeometry}Illustration of the co-flow geometry at the entrance of the co-flow region.}
\end{figure}

The RBC flow rate is measured by using a co-flow geometry \cite{guillot06,gachelin13}.  A microfluidics circuit is built through standard soft lithography techniques.  The geometry of the co-flow channel of thickness 40 \textmu m is illustrated in Fig. S\ref{figFlowGeometry}. Two inlet channels (width 500 \textmu m) are connected to a central co-flow channel (width $W_{tot}=$ 1000 \textmu m, length  1cm). In one of the inlet channels, an iso-dense solution is injected at flow rates $Q_{newt}$ imposed by a syringe pump ; in the other channel, RBC suspensions are injected at flow rates $Q_{T}$ controlled by another syringe pump.

In this co-flow geometry with a large aspect ratio, the pressure drop can be written as $\frac{\Delta P}{L} \propto \frac{Q_{i}\ \eta_{i}}{W_{i}}$, where the subscript $i$ refers either to one of the two fluids (RBC suspension or Newtonian iso-dense solution) and $W_i$ is the width occupied by fluid $i$. The pressure drop is kept constant over the course of an experimental run by keeping the flow rate $Q_{\mbox{newt}}$ constant and by adjusting the flow rate of the RBC suspensions $Q_{T}$ to keep the widths $W_i$  constant for all samples of different hematocrits. For the first sample, $Q_{\mbox{newt}}$  and $Q_{T}$  are chosen to provide a low enough velocity for the RBC suspension (of the order of mm $\cdot$ s$^{-1}$, $Q_{\mbox{newt}}$ between 1 and 3 $\mu$L/min.) and a width ratio $W_{newt}/W_{RBC}$ close to 1. This procedure allows us to keep the pressure drop constant along an experimental run without measuring it directly, by simple optical determination of the flow widths. $Q_T$ is then given by the value that was set on the syringe pump. It also ensures that the reference flow rate $Q_0$ (flow without cells) is equal to $Q_{\mbox{newt}}$.

\begin{figure}[h]
\centering
\includegraphics[width=9cm]{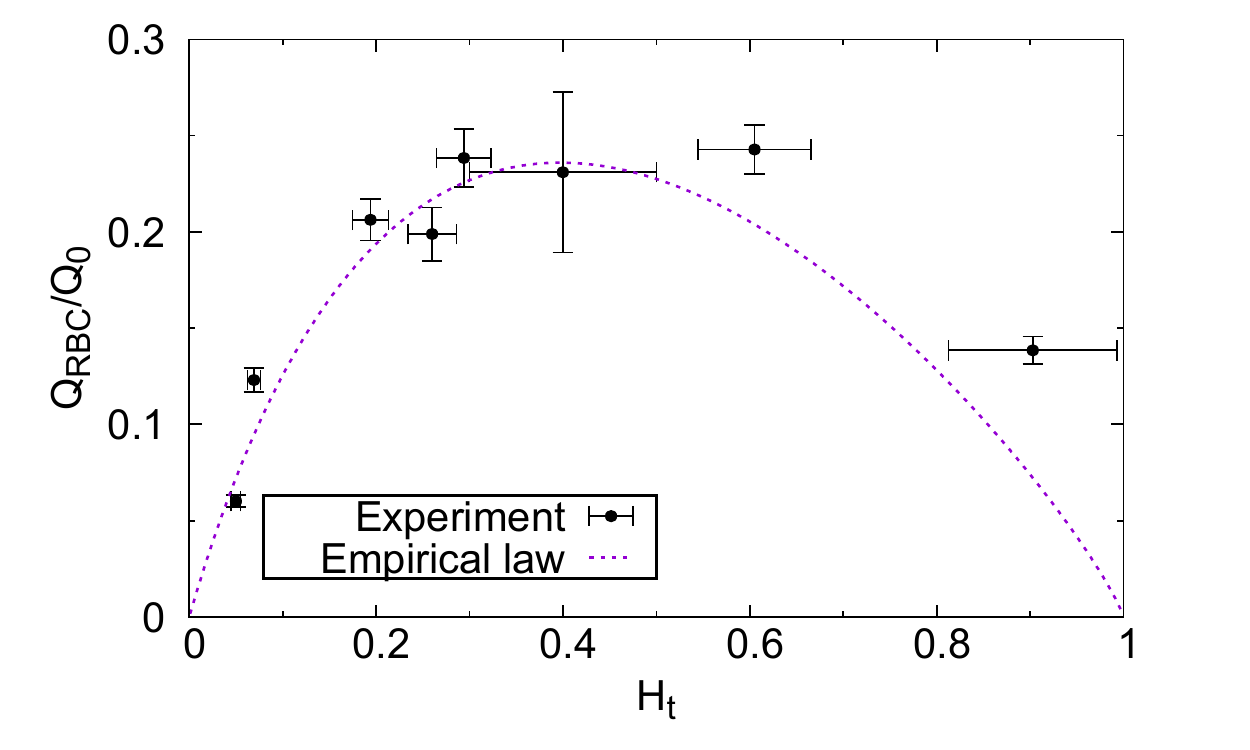}
\caption{\label{figresulexp} Reduced RBC flow rate as a function of tube hematocrit in a flat channel of thickness 40 \textmu m. The dashed line is calculated using the empirical Pries model.}
\end{figure}

The measured quantities $H_r$, $H_t$ and $Q_T$ are then combined to plot $Q_{RBC}/Q_0=H_r Q_T/Q_{\mbox{newt}}$ as a function of $H_t$, as in  Fig. S\ref{figresulexp}.

\section{Pries empirical model}

Pries \textit{et al.}  considered in \cite{Pries1992} a fit of experimental data to propose empirical relationship between the effective viscosity of blood in a cylindrical tube of diameter $D$ and the tube hematocrit $H_t$, as well as between this tube hematocrit and the reservoir hematocrit $H_r$.  Both functions $\eta_{\mbox{p}}(H_t)$ and $H_r(H_t)$  are given in  \cite{Pries1992}. The relative RBC flow rate at constant pressure drop is then given by $Q_{RBC}/Q_0=H_r(H_t)\times \frac{\eta_{\mbox{p}}(H_t=0)}{\eta_{\mbox{p}}(H_t)}$.

\section{Minimal model}

We derive more data from the minimal model used to identify the minimal ingredients necessary to understand the dependency of the RBC flow rate on the hematocrit, the channel size, but also with the RBC properties.

We consider a cylindrical channel of  radius $R_0=W/2$ with an outer ring of thickness $e$ where a fluid of viscosity 1 flows. In the core flows an homogeneous suspension of concentration $H$ and viscosity $\eta(H)=(1-H/H_m)^{-[\eta] H_m}$, where $H_m$ is the maximum volume fraction and $[\eta]$ the intrinsic viscosity.

As discussed in the main paper, $e$ is, in a first approximation, independent from $W$ and decreases linearly with increasing hematocrit. Self-consistency imposes then that    $e=e_0(1-H_t/H_m)$ , where $H_t= H \times (\frac{R_0-e}{R_0})^2$ is the tube hematocrit.
$Q_{RBC}/Q_0$  as a function of the tube hematocrit is given in the main paper.



\begin{figure}[h]
\centering
\includegraphics[width=8cm]{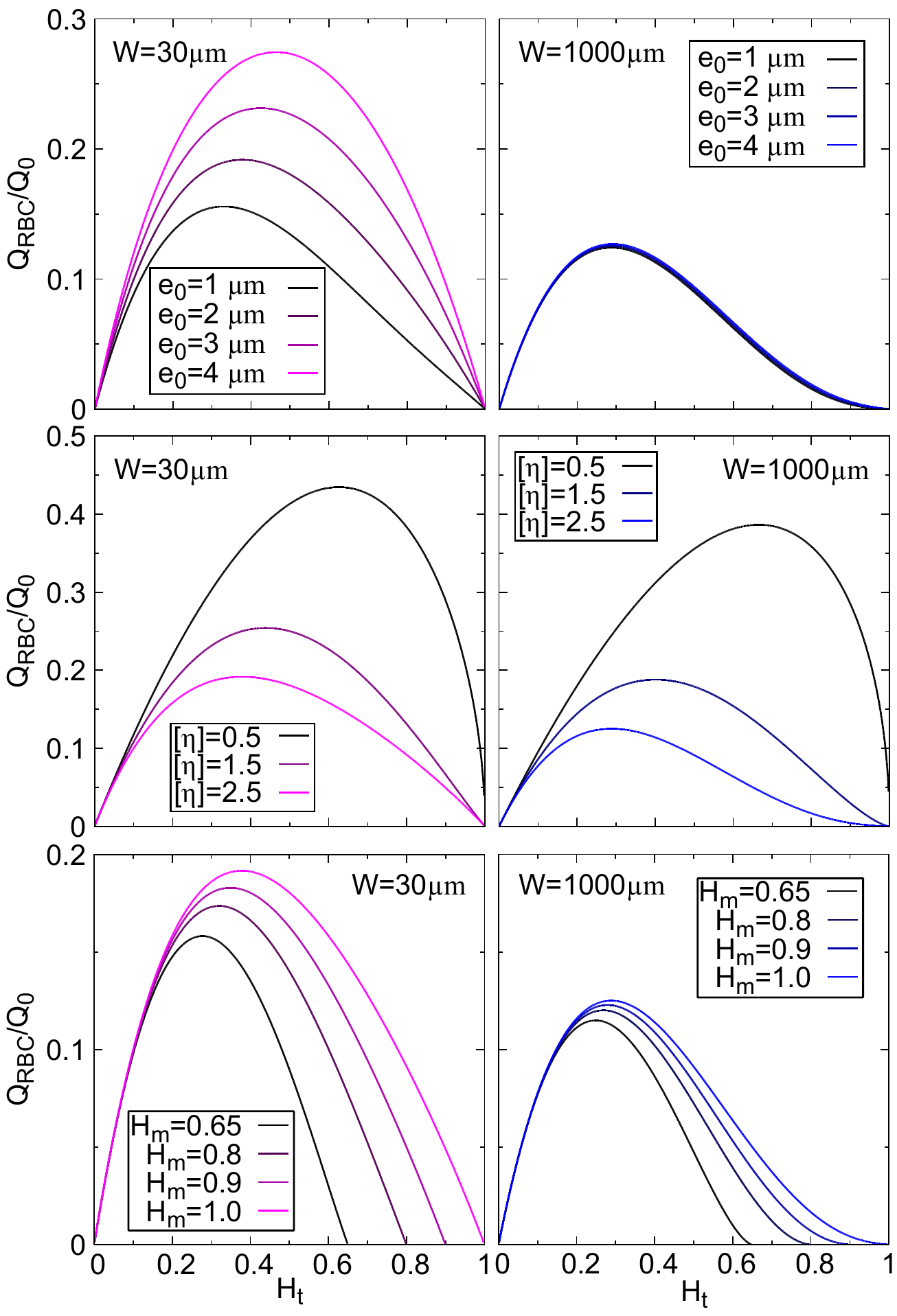}
\caption{\label{model} RBC flux according to our minimal model. The reference configuration is $W=30$ \textmu m (left) or  $W=1000$ \textmu m (right) , $e_0=2$ \textmu m, $[\eta]=5/2$ and $H_m=1$. Top panel: $e_0$ is varied ; middle panel: $[\eta]$ is varied ; bottom panel: $H_m$ is varied.}
\end{figure}

Here we exploit further the results of this simplified model to help proposing scenarios on the basis of our minimal model for the evolution of the optimal hematocrit with some cell mechanical properties: see Fig. S\ref{model}. For considering more deformable cells, a configuration which is classically explored by increasing the viscosity of the suspending fluid \cite{vitkova08,grandchamp13, shen16,dupire12,fischer13}, one expects the size of the cell free layer to increase \cite{shen16}. Also, the intrinsic viscosity would generally decrease, apart around the transition between tumbling and tank-treading regime,  as shown in \cite{vitkova08}. As seen  in Fig. S\ref{model}, both effects lead to an increase of the optimal hematocrit (except for large vessels for which the change in cell free layer size has no effect).

As discussed in the main paper, the evolution of the optimal hematocrit with the reduced volume can be understood by using our minimal model. If one now considers less deflated cells, the intrinsic viscosity is not expected to vary that much since it is equal to 5/2 for spheres and close to that value for RBCs. On the other hand, inflating cells would lead to a decrease of the maximal packing fraction $H_m$, which leads  to a decrease of the optimal hematocrit. Inflating cells would also lead to a decrease of the size of the cell free layer, which also leads to a decrease of the optimal hematocrit and of the associated RBC flux.

\section{Optimal hematocrit in vascular network}
We consider a simplified network consisting of $N$ levels enumerated as $i=1...N$. Level 1 consists of one straight cylindrical channel. Each channel of level $i$ has radius $R(i)$ and length $L(i).$ It divides symmetrically into two channels of level $i+1$. At each step, the RBC and plasma fluxes also split symmetrically, so that $H_r$ is a conserved quantity. We consider a fixed pressure drop $\Delta P$ between the inlet of the level-1 channel and the outlets of the level-$N$ channels.  Assuming  bifurcation junctions do not contribute to pressure drop, one straightforwardly obtains:
  \begin{equation} Q_{RBC} =H_r Q_T=H_r \Delta P \times\big\{\sum_{i=1}^N 2^{-i+1}\frac{8 L(i) \eta(H_r,R(i))}{\pi R(i)^4}\big\}^{-1}\label{eqnetwork}.\end{equation}

It is seen that the contribution of each channel depends on the detailed evolution of lengths and radii along the network. Recently, the relationships between radii $R(i)$ and lengths $L(i)$ currently proposed in the literature have been
  discussed in detail using a statistical analysis based on 3D imaging in human subjects \cite{newberry15}. Adapted to our case of dichotomous and symmetric branching, these relationships can be written as $R(i)^{1/a}=2 R(i+1)^{1/a}$ and $L(i)^{1/b}=2 L(i+1)^{1/b}$, where $a$ and $b$ are exponents that can be extracted from fitting of in-vivo data.
Using the recursive formulae $R(i)=2^{-a(i-1)}R(1)$ and $L(i)=2^{-b(i-1)}L(1)$ we find:
  \begin{equation}Q_{RBC} =H_r \frac{\pi R(1)^4 \Delta P}{8 L(1)} \times\big\{\sum_{i=1}^N 2^{(4a-b-1)(i-1)}\eta(H_r,R(i))\big\}^{-1}\label{eqnetworkscaled}.\end{equation}

The factor $2^{(4a-b-1)(i-1)}$ thus expresses the relative contribution from level $i$. According to \cite{newberry15}, a value for $a$ in the range $1/3-1/2$ yields a good description of real data, with $a=1/3$ for small vessels ($R<1$ mm) and $a=1/2$ for large vessels. An exponent-based scaling for the lengths is not as strongly supported by the data, since  different measurements yield  exponent $b$ in the range $0.17-1.40$. However, it is interesting to observe that for $a$ and $b$ within the aforementioned ranges, the exponent $4a-b-1$ can be either negative or positive, indicating that the major contributions can come from large or from small vessels depending on the exponent sign. Since the effective viscosity significantly depends on the tube radius only when the radius is smaller than about 1 mm \cite{pries92}, we focus on the case $a=1/3$, which is also consistent with  the  Murray's law \cite{Murray1926}. In that case, a realistic network geometry is likely to fall within a category where  $4a-b-1=1/3-b<0$ (since this is consistent with most available data for $b$). This result
 implies that  more weight is attributed to  wide vessels, yielding thus an optimal hematocrit $H_0$ around  known physiological values (see Fig. 7B in the main paper). Interestingly, this would point to the fact that optimization of oxygen transport capacity may be regulated  upstream of the capillaries, where
  blood gas exchanges with tissues and organs takes place.  However, given the dispersion of experimental data, the possibility for a positive exponent $4a-b-1=1/3-b$ is not to be excluded, in which case narrow channels  would have more weight than larger vessels. A systematic
analysis of the vessel size distribution in organs is needed before drawing more conclusive answers.

\end{document}